\documentclass[aps,article,showpacs,showkeys]{revtex4}%
\usepackage{amsfonts}
\usepackage{amsmath}
\usepackage{amssymb}
\usepackage{graphicx}%
\begin{document}
\title{Electromagnetic couplings of elementary vector particles}
\author{M.\ Napsuciale$^{(1,2)}$, S.\ Rodr\'{\i}guez$^{(3)}$, 
E.\ G.\ Delgado-Acosta$^{(1)}$, M.\ Kirchbach$^{(4)}$ }
\affiliation{$^{1}$ Instituto de F\'{\i}sica, Universidad de Guanajuato, 
Lomas del Bosque 103, Fraccionamiento Lomas del Campestre, 
37150, Le\'{o}n, Guanajuato, M\'{e}xico.}
\affiliation{$^{2}$ Departamento de F\'{\i}sica Te\'{o}rica and 
Instituto de F\'{\i}sica Corpuscular, Centro Mixto Universidad 
de Valencia-CSIC, 46000 Burjassot, Valencia, Spain.}
\affiliation{$^{3}$ Facultad de Ciencias F\'{\i}sico Matem\'{a}ticas, 
Universidad Aut\`{o}noma de Coahuila, Edificio "D", 
Unidad Camporredondo, CP 25280 Saltillo Coahuila.}
\affiliation{$^4$Instituto de F\'{\i}sica, Universidad Aut\'onoma 
de San Luis Potos\'{\i}, 
Av. Manuel Nava 6, San Luis Potos\'{\i}, S.L.P. 78290, M\'exico.}

\begin{abstract}
On the basis of the three fundamental principles of (i) Poincar\'{e} symmetry
of space time, (ii) electromagnetic gauge symmetry, and (iii) unitarity, we
construct an universal Lagrangian for the electromagnetic interactions of
elementary vector particles, i.e., massive spin-$1$ particles transforming in the
$\left( \frac{1}{2},\frac{1}{2}\right)$ representation space
of the Homogeneous Lorentz Group (HLG). 
We make the point that the first two 
symmetries alone do not fix the electromagnetic couplings uniquely but
solely prescribe a general Lagrangian depending on two free 
parameters, here denoted by $\xi$ and $g$. 
The first one defines the electric-dipole and the
magnetic-quadrupole moments of the vector particle, while the second 
determines its  magnetic-dipole and  electric-quadrupole moments. 
In order to fix the parameters one needs an additional physical 
input suited for the implementation of the third principle.
As such, one chooses Compton scattering off a vector target  
and  requires the cross section to respect the 
unitarity bounds in the high energy limit.
In result, we obtain the universal $g=2$, and $\xi=0$ values which
completely characterize the electromagnetic couplings of the considered
elementary vector field at tree level. 
The nature of this vector particle, Abelian
versus non-Abelian, does not affect this structure. Merely, a partition of the
$g=2$ value into non-Abelian, $g_{na}$, and Abelian, $g_{a}=2-g_{na}$,
contributions occurs for non-Abelian fields with the size of $g_{na}$ being
determined by the specific non-Abelian group appearing in the theory of
interest, be it the Standard Model or any other theory.

\end{abstract}
\maketitle

\section{Introduction}

In the forthcoming years, energies ranging from several hundreds GeV to few
TeV's are expected to become accessible at the particle accelerators, a
progress which will facilitate testing various fundamental theoretical
concepts. In particular, it is quite possible that some of the elementary
high-spin particles predicted by supersymmetric-, or, excited-lepton theories
could be observed either as gauge fields to some still unknown non-Abelian
groups, or, as matter fields. Additional effects may or may not come from the
more recently developed respective theories of large extra dimensions,
non-commutative space-time etc. In view of the theoretical uncertainties it
appears quite important indeed to single out the impact of the first
principles underlying the space-time on the properties of the elementary
high-spin fields and in first place on their electromagnetic properties. As an
example one may think of the value of the gyromagnetic factor, $g$. In recent
time, voicing universality of the $g=2$ value for particles of any spin
becomes stronger (see ref.~\cite{BarryH} and references therein for a recent
review). An indirect indication in favor of $g=2$ is provided already by the
Drell-Hearn-Gerasimov sum rule \cite{Drell-Hearn} (generalized by Weinberg
\cite{Weinberg} to any spin) which assigns to strong interactions the
anomalous magnetic moment of the nucleon in terms of the $(g-2)e/2m$
difference. Ferrara, Porrati and Telegdy made the point \cite{Ferr} that
unitarity of the amplitude of Compton scattering off a target of \textit{any
spin\/} $s$ demands for the universal value of $g=2$. Specifically for
spin-$3/2$, they showed that $g=2$ also allows to avoid the pathology of
acausal propagation \cite{VZ32} of such particles within an electromagnetic
environment as suffered by the Rarita-Schwinger formalism. However, the
resolution of this so called Velo-Zwanziger problem was obtained at the cost
of the introduction of \textit{non-minimal \/} electromagnetic couplings.
In contrast to the Ferrara, Porrati and Telegdy approach, in the recently
proposed covariant projector formalism of ref.~\cite{NKR}, spin-$3/2$ causal
propagation and $g=2$ were achieved by means of a Lagrangian containing
\textit{only minimal\/} couplings but of \textit{second order\/} 
in the momenta. The latter formalism treats high-spins as appropriate 
sectors of finite-dimensional multi-spin valued HLG representations 
that behave as invariant eigensubspaces of the two Casimir operators of 
the Poincar\'e group, the squared four momentum, $p^{2}$, and the 
squared Pauli-Lubanski vector,
$\mathcal{W}^{2}$. It is the goal of the present study to apply the
covariant projector formalism to fields residing in the
$\left(\frac{1}{2},\frac{1}{2}\right)$ irreducible representation of the HLG,
the massive vector fields, and explore consequences on their electromagnetic couplings.

The paper is organized as follows. In the following section we briefly review
the covariant projector framework of ref.~\cite{NKR} with the emphasis on
the $\left( \frac{1}{2},\frac{1}{2}\right)$ representation of the HLG 
and work out the electromagnetic interactions. In section III we
calculate the cross section for Compton scattering off a vector target. In
section IV we discuss our results within the light of the Abelian and
non-Abelian contributions to the tree level electromagnetic couplings of
arbitrary gauge fields. The paper ends with brief conclusions and has 
one Appendix.

\section{Description of vector particles}

\subsection{General remarks on vector fields}

 Vector fields, $V_{\mu}$, have been previously studied by several
authors with the emphasis on their electromagnetic properties. Recently,
it was pointed out in \cite{Toledo} that $g=2$ is required to avoid appearance
of divergent $\mathcal{O}(\omega^{-1})$ terms in the radiative decay
interferences for polarized vector mesons, with $\omega$ standing for the
photon energy. On the other side, Proca's theory goes with a fixed $g=1$ value
and it is not very clear how to reconcile it with $g=2$ except for
the $W$ boson, the gauge particle of the electroweak
$SU(2)_{L}\times U(1)_{Y}$ group.

\noindent The construction of the interacting $(W^{+}W^{-}\gamma)$ Lagrangian is
quite intricate indeed. The minimally gauged Proca Lagrangian is complemented
by a Lagrangian of the same Proca form but based on the non-Abelian field tensor
\cite{BarryH}. The contribution of $g=1$ of the former is then enhanced
precisely by the required one unit through the latter after $SU(2)_{L}\times
U(1)_{Y}/U_{em}(1)$ spontaneous symmetry breaking, to give $g=2$ (see
ref.~\cite{Hey} for a textbook presentation). Here, $U_{em}(1)$ stands for the
electromagnetic gauge group. In this manner, the gyromagnetic ratio of the $W$
boson is equally partitioned into Abelian and non-Abelian contributions. Such
a symmetrical partition is not likely to be universal, although in the special
case of the $W$ boson some unification theories seem to preserve it
\cite{Lee-Kim}. However, for different vector gauge bosons, the new
non-Abelian theories throughout may provide larger or lesser non-Abelian
contributions to $g$. Within this context, it is desirable to have a scheme 
for the description of vector fields that goes beyond Proca's formalism and 
allows to end up with a $g=2$ for any vector particle irrespective of its 
nature, Abelian or non-Abelian.

\begin{quote}
In the present work we derive such a scheme and prove that the electromagnetic
couplings at tree level of \textit{any} massive elementary vector particle are
completely fixed by the three fundamental principles of (i) Poincar\'{e}
symmetry of space time, (ii) $U(1)_{em}$ gauge symmetry, and (iii) unitarity.
Modifications to this picture can arise only at one loop level due to 
electromagnetic corrections or diagrams involving interactions with 
other fields.
\end{quote}

We first make the point that the Proca framework is incomplete in observing
that the Proca Lagrangian neglects viable terms containing anti-commutators,
$\lbrack p_{\mu},p_{\nu}\rbrack$, of the four-momenta, which do not contribute
to the free equation of motion at all, but affect the electromagnetic moments
in the gauged one when they become proportional to the electromagnetic field
tensor, $F_{\mu\nu}$, according to $\lbrack\pi_{\mu},\pi_{\nu}\rbrack
=ieF_{\mu\nu} $ with $\pi_{\mu}=p_{\mu}+eA_{\mu}$. We will show below that the
unique $g=1$ value in the Proca theory upon $U(1)_{em}$ gauging appears
precisely as an artifact of the mentioned shortcoming of the free Proca
Lagrangian. This shortcoming has been avoided within the framework of the
covariant projector formalism recently suggested in ref.~\cite{NKR}. Within
this context, exploring the electromagnetic properties of vector particles
within the latter scheme is worthwhile.

In the following we shall obtain a general Lagrangian for a vector
particle whose interaction  with an electromagnetic field 
is consistent with Poincar\'e symmetry as 
implemented by the covariant projection formalism of 
ref.~ \cite{NKR}, and $U(1)_{em}$ gauge principle. In contrast to
Proca's framework, we shall encounter not one but infinitely many equivalent
free particle theories which, upon gauging, begin differing through their
predictions on the values of the multipole moments, only one of which
corresponds to physical reality. In order to fix these values, one then needs
additional physical input. As such we consider Compton scattering off a vector 
target and demand finite total cross section in the high-energy limit in 
order to respect the unitarity bounds. Taking this path allows us to 
completely fix the electromagnetic couplings of any elementary vector particle at tree level. 
In fact, there is no freedom left any more in the Lagrangian designed to account for
all possible terms containing $\lbrack p _{\mu},p_{\nu}\rbrack$
anti-commutators, terms that are notoriously missed by the Proca theory.

\subsection{The covariant projector formalism}

In ref.~\cite{NKR}, a formalism was proposed which describes fields of mass
$m$ and spin $s$ from a given finite dimensional and multi-spin valued
representation of the homogeneous Lorentz group (HLG) in terms of simultaneous
projection over the eigensubspaces of the two Casimir operators of the
Poincar\'{e} group, the squared four-momentum, $p^{2}$, and the squared
Pauli-Lubanski vector $\mathcal{W}^{2}$. In particular, in the case of
representations containing two different spin-values differing in one
unit, say, $s$ and $(s-1)$, the free particle equation obtained in this way
reads
\begin{equation}
-\frac{p^{2}}{m^{2}}\frac{1}{2s}\left[  \frac{{\mathcal{W}}^{2}}{p^{2}%
}+s(s-1)\openone\right]  _{AB}\Psi_{B}^{(m,s)}=\Psi_{A}^{(m,s)}, \label{PML}%
\end{equation}
where capital Latin letters $A,B,C,...$ specify the HLG representation of
interest. The general expression for the $\mathcal{W}^{2}$ operator can be
found in \cite{NKR} and reads
\begin{equation}
(\mathcal{W}_{\lambda}\mathcal{W}^{\lambda})_{AB}=\frac{1}{4}\epsilon
_{\lambda\rho\sigma\mu}(M^{\rho\sigma})_{AC}p^{\mu}\epsilon_{~\tau\xi\nu
}^{\lambda}(M^{\tau\xi})_{CB}p^{\nu}\equiv T_{AB\mu\nu}p^{\mu}p^{\nu}.
\label{W2}%
\end{equation}
From now onward we shall introduce as a new notation the tensor
$\widetilde{\Gamma}_{AB\mu\nu}$ according to
\begin{equation}
\widetilde{\Gamma}_{AB\mu\nu}=-\frac{1}{2s}\left(  T_{AB\mu\nu}+s(s-1)\delta
_{AB}\ g_{\mu\nu}\right)  .
\end{equation}
Notice, that for the $\left(  \frac{1}{2},\frac{1}{2}\right)  $ representation
space, the capital Latin indices coincide with the 
Lorentz indices. A straightforward calculation
(see Appendix I in \cite{NKR}) yields
\begin{equation}
\widetilde{\Gamma}_{\alpha\beta\mu\nu}=g_{\alpha\beta}g_{\mu\nu}-g_{\alpha\nu
}g_{\beta\mu}. \label{Gamma}%
\end{equation}
Using this tensor, the free equation of motion for a vector particle becomes%
\begin{equation}
\left[  \widetilde{\Gamma}_{\alpha\beta\mu\nu}\partial^{\mu}\partial^{\nu
}+m^{2}g_{\mu\nu}\right]  V^{\nu}=0, \label{eom}%
\end{equation}
where $V^{\nu}$ denotes the vector field. The tensor in eq.~(\ref{Gamma})
can be decomposed into its symmetric and antisymmetric parts as
\begin{equation}
\widetilde{\Gamma}_{\alpha\beta\mu\nu}=\widetilde{\Gamma}_{\alpha\beta\mu\nu
}^{S}+\widetilde{\Gamma}_{\alpha\beta\mu\nu}^{A},
\end{equation}
with%
\begin{align}
\widetilde{\Gamma}_{\alpha\beta\mu\nu}^{S}  &  =\frac{1}{2}\left(
\widetilde{\Gamma}_{\alpha\beta\mu\nu}+\widetilde{\Gamma}_{\alpha\beta\nu\mu
}\right)  =g_{\alpha\beta}g_{\mu\nu}-\frac{1}{2}(g_{\alpha\nu}g_{\beta\mu
}+g_{\alpha\mu}g_{\beta\nu}),\label{sym}\\
\widetilde{\Gamma}_{\alpha\beta\mu\nu}^{A}  &  =\frac{1}{2}\left(
\widetilde{\Gamma}_{\alpha\beta\mu\nu}-\widetilde{\Gamma}_{\alpha\beta\nu\mu
}\right)  =\frac{1}{2}(g_{\alpha\mu}g_{\beta\nu}-g_{\alpha\nu}g_{\beta\mu}).
\label{asym}%
\end{align}
As discussed in \cite{NKR} and is also evident from eq.~(\ref{eom}), the
covariant mass-and spin projector in eq.~(\ref{PML}) fixes uniquely only the
part of the tensor $\Gamma_{\alpha\beta\mu\nu}$ that is symmetric in the
indices $(\mu,\nu)$. Equation~(\ref{eom}) is indisputably insensitive to the
antisymmetric part which acquires relevance exclusively upon gauging when
$\lbrack p_{\mu}, p_{\nu}\rbrack$ become proportional to the electromagnetic
field tensor, $F_{\mu\nu}$, according to $\lbrack\pi_{\mu},\pi_{\nu}%
\rbrack=ieF_{\mu\nu} $ with $\pi_{\mu}=p_{\mu}+eA_{\mu}$. Indeed, it is
precisely the $\widetilde{\Gamma}_{\alpha\beta\mu\nu}^{A}$ term which triggers
the interactions with multipoles higher than the electric charge. A complete
formalism requires to account for the most general form of the 
antisymmetric tensor.

\subsection{General Lagrangian for an elementary vector particle in an electromagnetic background.}

In the vector case under investigation, the most general tensor $\Gamma_{\alpha\beta
\mu\nu}^{A}$ has to be constructed from the metric-- and
the Levi-Civita tensors and is given by%
\begin{equation}
\Gamma_{\alpha\beta\mu\nu}^{A}=\left(  g-\frac{1}{2}\right)  (g_{\alpha\mu
}g_{\beta\nu}-g_{\alpha\nu}g_{\beta\mu})+\xi\text{ }\varepsilon_{\alpha
\beta\mu\nu}, \label{AS_general}%
\end{equation}
where $g$ and $\xi$ are free parameters, so far. In what follows we shall
replace $\widetilde{\Gamma}_{\alpha\beta\mu\nu}^{A}$ in eq.~(\ref{asym}) by
$\Gamma_{\alpha\beta\mu\nu}^{A}$ from the last equation. In result, the most
general tensor compatible with the covariant mass--and spin projector in
eq.~(\ref{PML}) becomes
\begin{equation}
\Gamma_{\alpha\beta\mu\nu}=g_{\alpha\beta}g_{\mu\nu}+\left(  g-1\right)
g_{\alpha\mu}g_{\beta\nu}-g\ g_{\alpha\nu}g_{\beta\mu}+\xi\text{ }%
\varepsilon_{\alpha\beta\mu\nu}. \label{Gammag}%
\end{equation}
The corresponding gauged equation of motion is then obtained as
\begin{equation}
\left[  \Gamma_{\alpha\beta\mu\nu}D^{\mu}D^{\nu}+m^{2}g_{\mu\nu}\right]
V^{\nu}=0. \label{eomg}%
\end{equation}
It can be derived from the following Lagrangian%
\begin{equation}
\mathcal{L}=-(D^{\mu}V^{\alpha})^{\dagger}\Gamma_{\alpha\beta\mu\nu}D^{\nu
}V^{\beta}+m^{2}V^{\alpha}V_{\alpha}, \label{Lagg}%
\end{equation}
where $D^{\mu}=\partial^{\mu}-ieA^{\mu}$ and $(-e)$ is the charge of the
vector particle. The hermiticity of the Lagrangian requires the couplings $g$,
$\xi$ to be real. Although the projection over the eigensubspaces of the
Casimir operators of the Poincar\'{e} group studied here is well defined for
massive particles only, the free Lagrangian
\begin{equation}
\mathcal{L}_{free}=-(\partial^{\mu}V^{\alpha})^{\dagger}\Gamma_{\alpha\beta
\mu\nu}\partial^{\nu}V^{\beta}+m^{2}V^{\alpha}V_{\alpha}, \label{Lag}%
\end{equation}
possesses a smooth massless limit. In this limit the free Lagrangian reveals
as a symmetry the invariance under the $U_{V}(1)$ gauge transformations%
\begin{equation}
V_{\alpha}\rightarrow V_{\alpha}+\partial_{\alpha}\Lambda.
\end{equation}
The mass term can now be generated through the conventional Higgs mechanism
\cite{NKR} in reference to this symmetry. A straightforward calculation yields
the following interacting Lagrangian
\begin{equation}
\mathcal{L}_{int}=-ie[\left(  V^{\alpha}\right)  ^{\dagger}\Gamma_{\alpha
\beta\mu\nu}\partial^{\nu}V^{\beta}-(\partial^{\nu}V^{\alpha})^{\dagger}%
\Gamma_{\alpha\beta\nu\mu}V^{\beta}]A^{\mu}+e^{2}\left(  V^{\alpha}\right)
^{\dagger}\Gamma_{\alpha\beta\mu\nu}V^{\beta}A^{\mu}A^{\nu}. \label{Lint}%
\end{equation}
The respective $V^{\beta}(p)V^{\alpha}(p^{\prime})A^{\mu}(k)$, and $V^{\alpha
}(p^{\prime})V^{\beta}(p)A^{\mu}(k)A^{\nu}(k^{\prime})$ vertex functions
extracted from eq.~(\ref{Lint}) read%
\begin{equation}
V_{a\beta\mu}=ie(\Gamma_{\alpha\beta\mu\nu}p^{\nu}-\Gamma_{\alpha\beta\nu\mu
}p^{\prime\nu}),\qquad V_{a\beta\mu\nu}=-ie^{2}(\Gamma_{\alpha\beta\mu\nu
}+\Gamma_{\alpha\beta\nu\mu}),
\end{equation}
with all incoming particles. Explicitly%
\begin{equation}
V_{a\beta\mu}=ie\left(  g_{\alpha\beta}\left(  p-p^{\prime}\right)  _{\mu
}-g_{\alpha\mu}\left[  g\ k+p\right]  _{\beta}+g_{\beta\mu}\left[  p^{\prime
}+g\ k\right]  _{\alpha}+\xi\text{ }\varepsilon_{\alpha\beta\mu\nu}\left(
p+p^{\prime}\right)  ^{\nu}\right)  . \label{VNKR}%
\end{equation}
This vertex describes the electromagnetic interactions of a particle with
magnetic (electric) dipole moment $\mu$ ($\widetilde{\mu}$) and quadrupole
electric (magnetic) moment $Q$ ($\widetilde{Q}$) given by ( see e.g.
ref.(\cite{Toscano} )
\begin{equation}
\mu=\frac{ge}{2m},\qquad Q=-\frac{\left(  g-1\right)  e}{m^{2}},\qquad
\widetilde{\mu}=\frac{\xi e}{2m},\qquad\widetilde{Q}=-\frac{\xi e}{m^{2}}.
\end{equation}
It satisfies the Ward identity
\begin{equation}
(p+p^{\prime})^{\mu}V_{a\beta\mu}=-ie\left[  \Delta_{\alpha\beta}%
^{-1}(p^{\prime})-\Delta_{\alpha\beta}^{-1}(p)\right]  , \label{Ward}%
\end{equation}
where $\Delta_{\alpha\beta}(p)$ is the propagator of the massive vector
particle which in the unitary gauge (with respect to the gauge freedom in the
massless case, see \cite{NKR}) we are using here reads
\begin{equation}
\Delta_{\alpha\beta}\left(  p\right)  =\frac{-g_{\alpha\beta}+\frac{p_{\alpha
}p_{\beta}}{m^{2}}}{p^{2}-m^{2}+i\varepsilon}.
\end{equation}
Notice that Poincar\'{e} and gauge invariance alone allow the vector 
particle to carry
any arbitrary magnetic and electric dipole moments which then enter the
definition of the electric and magnetic quadrupole moments, respectively. As
long as the electric dipole and the magnetic quadrupole moments are CP
violating, the $\xi$ value is expected to be rather small. Nonetheless, we
will keep this term for the sake of completeness of our Poincar\'{e} covariant
projector and will fix it from unitarity arguments together with $g$. The Proca theory
corresponds instead to a \textit{fixed unphysical} $g=1$ value and, in being
incomplete, as mentioned in the introduction, fails to predict a quadrupole
electric moment.

\noindent A Lagrangian for vector particles containing $g$ as a free parameter has earlier
been considered by Corben and Schwinger \cite{Corben-Schwinger} and used later
by Lee and Yang \cite{Lee-Yang}. In contrast to our approach, Poincar\'e
invariance is not made manifest in the Corben-Schwinger paper, but is somehow
hidden in the restriction of all derivatives to second order, and $\left(
\frac{1}{2},\frac{1}{2}\right)  $ to spin-$1$. In our formalism the
restriction of derivatives to second order is dictated by the squared
Pauli-Lubanski operator in eq.(\ref{W2}) around which the Poincar\'{e}
projector is constructed. Compared to \cite{Corben-Schwinger}, the advantage
of our scheme lies in its generality. Consciously putting first principles
at work, sheds light on the path for obtaining the most general Lagrangian
for a particle of spin $s$ transforming in a specific representation of the HLG. 
Moreover, in the next section we will add another
fundamental principle to the first two, namely unitarity, which will allow us
to completely fix the electromagnetic couplings of an elementary vector particle at tree 
level.

\noindent Our interacting Lagrangian with the unspecified value for
the gyromagnetic ratio and the electric dipole moment (and the related
quadrupoles as shown above) appeared as a consequence of the fact that
Poincar\'{e} invariance provides not one but infinitely many equivalent free
particle Lagrangians according to eq.~(\ref{Lagg}) in combination with
eq.~(\ref{Gammag}). These Lagrangians become distinguishable only upon gauging
precisely through the different values for the respective gyromagnetic ratio,
and electric dipole moment predicted by them. Obviously, only one of the $g
(\xi)$ possible values corresponds to physical reality. In order to fix these
values, one needs additional physical information. In the present work we
shall demand finite total cross section of Compton scattering off a vector
target in the high energy limit and determine $g$ and $\xi$ accordingly.

\section{Compton scattering off a vector target}

The differential cross section for $\gamma(k,\epsilon)V(p,\zeta)\rightarrow
\gamma(k^{\prime},\epsilon^{\prime})V(p^{\prime},\zeta^{\prime})$ in the
laboratory frame is given as
\begin{equation}
\frac{d\sigma}{d\Omega}=\frac{1}{4\left(  4\pi\right)  ^{2}}\left\vert
\mathcal{\bar{M}}\right\vert ^{2}\frac{1}{\left(  m+\omega(1-\cos
\theta)\right)  ^{2}},
\end{equation}
where $\omega$ stands for the energy of the incoming photon. The invariant
amplitude can be written as
\begin{equation}
\mathcal{M}=\mathcal{M}_{s}+\mathcal{M}_{u}+\mathcal{M}_{c},\label{amp}%
\end{equation}
where $\mathcal{M}_{s}$, $\mathcal{M}_{u}$ and $\mathcal{M}_{c}$ denote in
turn the contributions of the $s$-, and $u$- channel exchange, and the contact
term. The explicit forms of these amplitudes are%
\begin{align}
\mathcal{M}_{s} &  =\left[  V_{\sigma\beta\mu}\left(  p,-p-k\right)
\Delta^{\sigma\rho}\left(  p+k\right)  V_{\alpha\rho\nu}(p^{\prime}+k^{\prime
},-p^{\prime})\right]  \zeta^{\beta}\epsilon^{\mu}\zeta^{\prime\alpha}%
\epsilon^{\prime\nu}\,,\\
\mathcal{M}_{u} &  =\left[  V_{\sigma\beta\nu}\left(  p,-p+k^{\prime}\right)
\Delta^{\sigma\rho}\left(  p-k^{\prime}\right)  V_{\alpha\rho\mu}(p^{\prime
}-k^{\prime},-p^{\prime})\right]  \zeta^{\beta}\epsilon^{\mu}\zeta
^{\prime\alpha}\epsilon^{\prime\nu}\,,\\
\mathcal{M}_{c} &  =V_{a\nu\beta\mu}\zeta^{\beta}\epsilon^{\mu}\zeta^{\alpha
}\epsilon^{\nu}\,.
\end{align}
As a check, replacing $\epsilon^{\mu}$ by $k^{\mu}$ and using the Ward
identity we obtain%
\begin{align}
\mathcal{M}_{s}(\epsilon^{\mu}\rightarrow k^{\mu}) &  =e^{2}(\Gamma
_{\alpha\beta\nu\mu}\left(  p+k\right)  ^{\mu}+\Gamma_{\alpha\beta\mu\nu
}p^{\prime\mu})\zeta^{\beta}\zeta^{\prime\alpha}\epsilon^{\prime\nu},\\
\mathcal{M}_{u}(\epsilon^{\mu}\rightarrow k^{\mu}) &  =-e^{2}(\Gamma
_{\alpha\beta\nu\mu}p^{\mu}+\Gamma_{\alpha\beta\mu\nu}\left(  p^{\prime
}-k\right)  ^{\mu})\zeta^{\beta}\zeta^{\prime\alpha}\epsilon^{\prime\nu},\\
\mathcal{M}_{c}(\epsilon^{\mu}\rightarrow k^{\mu}) &  =V_{a\nu\beta\mu}%
\zeta^{\beta}\zeta^{\prime\alpha}k^{\mu}\epsilon^{\prime\nu}.
\end{align}
Upon summing up the three contributions one sees that gauge invariance is
satisfied
\begin{equation}
\mathcal{M}\left(  \epsilon\rightarrow k\right)  =0.
\end{equation}
A similar calculation for the outgoing photon confirms once again gauge
invariance to be satisfied. Using the conditions $k\cdot\varepsilon=k^{\prime
}\cdot\varepsilon^{\prime}=p\cdot\zeta=p^{\prime}\cdot\zeta^{\prime}=0$, 
we calculated $\mathcal{M}$ 
explicitly in eq.~(\ref{Amplitude}) in the appendix.
Inspection of the latter expression shows that
the divergent terms in the high energy limit come from the $1/m^{2}$ terms 
which are proportional to $(g-2)$, $\xi$, their 
product and their second power, thus, they vanish for $g=2$ and $\xi=0$.
Another and perhaps easier way to see that such cancellation occurs 
only for the mentioned values  is to calculate the cross section. 
A straightforward calculation of the squared amplitude yields 
eq.~(\ref{dsgt}) from the appendix.
The latter expression shows that in the classical limit,
$\eta\rightarrow 0$, the differential cross section is independent of $g$ and
$\xi$, as it should be,
\begin{equation}
\left.  \frac{d\sigma(g,\xi)}{d\Omega}\right\vert _{\eta\rightarrow0}%
=\frac{r_{0}^{2}}{2}\left(  1+x^{2}\right)  ,
\end{equation}
and the total cross section coincides with the Thompson result,%
\begin{equation}
\left.  \sigma(g,\xi)\right\vert _{\eta\rightarrow0}=\frac{8\pi}{3}r_{0}%
^{2}\equiv\sigma_{T}.
\end{equation}
More interesting is the high-energy limit, $\eta\gg1$, in which case we find
\begin{align}
\left.  \frac{d\sigma(g,\xi)}{d\Omega}\right\vert _{\eta\gg1} &  =\frac
{r_{0}^{2}}{96\,\left(  -1+x\right)  ^{2}}\left[  80+g^{4}\,\left(  21+\left(
-8+x\right)  \,x\right)  +8\,g^{3}\,\left(  -10+\left(  -1+x\right)
\,x\right)  +88\,\xi^{2}+21\,\xi^{4}-8\,x\,\xi^{4}\right.  \label{dgdsinfty}\\
&  +x^{2}\,{\left(  -4+\xi^{2}\right)  }^{2}+2\,g^{2}\,\left(  4\,\left(
17+x\,\left(  4+x\right)  \right)  +\left(  21+\left(  -8+x\right)
\,x\right)  \,\xi^{2}\right)  \nonumber\\
&  +\left.  8\,g\,\left(  -4\,\left(  4+x+x^{2}\right)  +\left(  -10+\left(
-1+x\right)  \,x\right)  \,\xi^{2}\right)  \right]  .\nonumber
\end{align}
In general, for arbitrary $g$ and $\xi$, the angular distribution of the
emitted photon is sharply peaked in forward direction and the total cross
section diverges violating the unitarity bounds
\cite{Ferr,Booth-Wilson,Bludman-Young}. In order to check the values of $g$
and $\xi$ avoiding this ultraviolet catastrophe we integrate the differential
cross section in eq. (\ref{dgdsinfty}) from $x=-1+\epsilon$ to $x=1-\epsilon$,
with $\epsilon\rightarrow0$ to obtain
\begin{align}
\left.  \sigma(g,\xi)\right\vert _{\eta\gg1} &  =\frac{8\pi r_{0}^{2}}{3}%
\frac{1}{128}\left[  2\left(  1-\epsilon\right)  \,\left(  g^{2}+4g-4+{\xi
}^{2}\right)  ^{2}\right.  \label{sinfty}\\
&  +2\,\left(  \left(  g-2\right)  ^{2}+{\xi}^{2}\right)  \,\left(
7\,g^{2}-12g+12+7\,{\xi}^{2}\right)  \left(  \frac{1}{\epsilon-2}+\frac
{1}{\epsilon}\right)  \nonumber\\
&  +\left.  2\,\left(  \left(  g-2\right)  ^{2}+{\xi}^{2}\right)  \,\left(
3\,g^{2}+8g-4+3\,{\xi}^{2}\right)  \,\log(\frac{2}{\epsilon}-1)\right]
.\nonumber
\end{align}

The latter equation makes manifest that the only values of $g$ and $\xi$
preventing the violation of unitarity at high energies are
indeed $g=2$, and $\xi=0$, respectively.

\begin{quote}
Using the first principles of (i) the covariant projection on the mass-$m$ and spin-$1$
eigensubspace of the Casimir operators of the Poincar\'{e} Group in the
$\left(\frac{1}{2},\frac{1}{2}\right)  $ representation space of the HLG
in combination with the most general form of
the anti-symmetric part of the corresponding tensor (not fixed by the
projection), (ii) $U(1)_{em}$ gauge principle, and (iii) unitarity in the
high-energy limit, we were able to uniquely fix the electromagnetic couplings
of \textit{any} elementary vector particle at tree level. Any massive spin-1 particle
described by means of a four-vector field must have a magnetic dipole moment of 
$\mu=e/m$, an electric quadrupole moment
of $Q=-e/m^{2}$, and vanishing electric dipole and magnetic quadrupole moments
at tree level. This is the prime result of this work.
\end{quote}

\begin{figure}[ptb]
\begin{center}
\includegraphics[
natheight=6.0in,
natwidth=10.0in,
height=4.0in,
width=6.0in
]{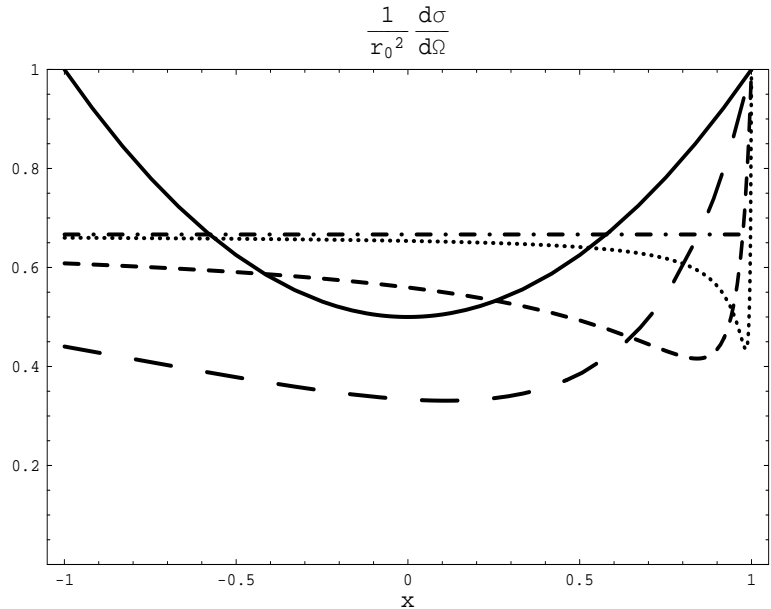}
\end{center}
\caption{Differential cross section normalized to the classical squared radius
as a function of $x=\cos\theta$ for different values of the energy of the
incident photon in the laboratory frame: $\eta=0$ (thick line), $\eta=1$ (long
dashed ) $\eta=10$ (short-dashed), $\eta=100$ (dotted) and $\eta=\infty$
(dot-dashed), where  $\eta=\omega/m$.}%
\label{dsig}%
\end{figure}

\begin{center}
\begin{figure}[ptb]
\begin{center}
\includegraphics[
natheight=6.0in,
natwidth=10.0in,
height=4.0in,
width=6.5in
]{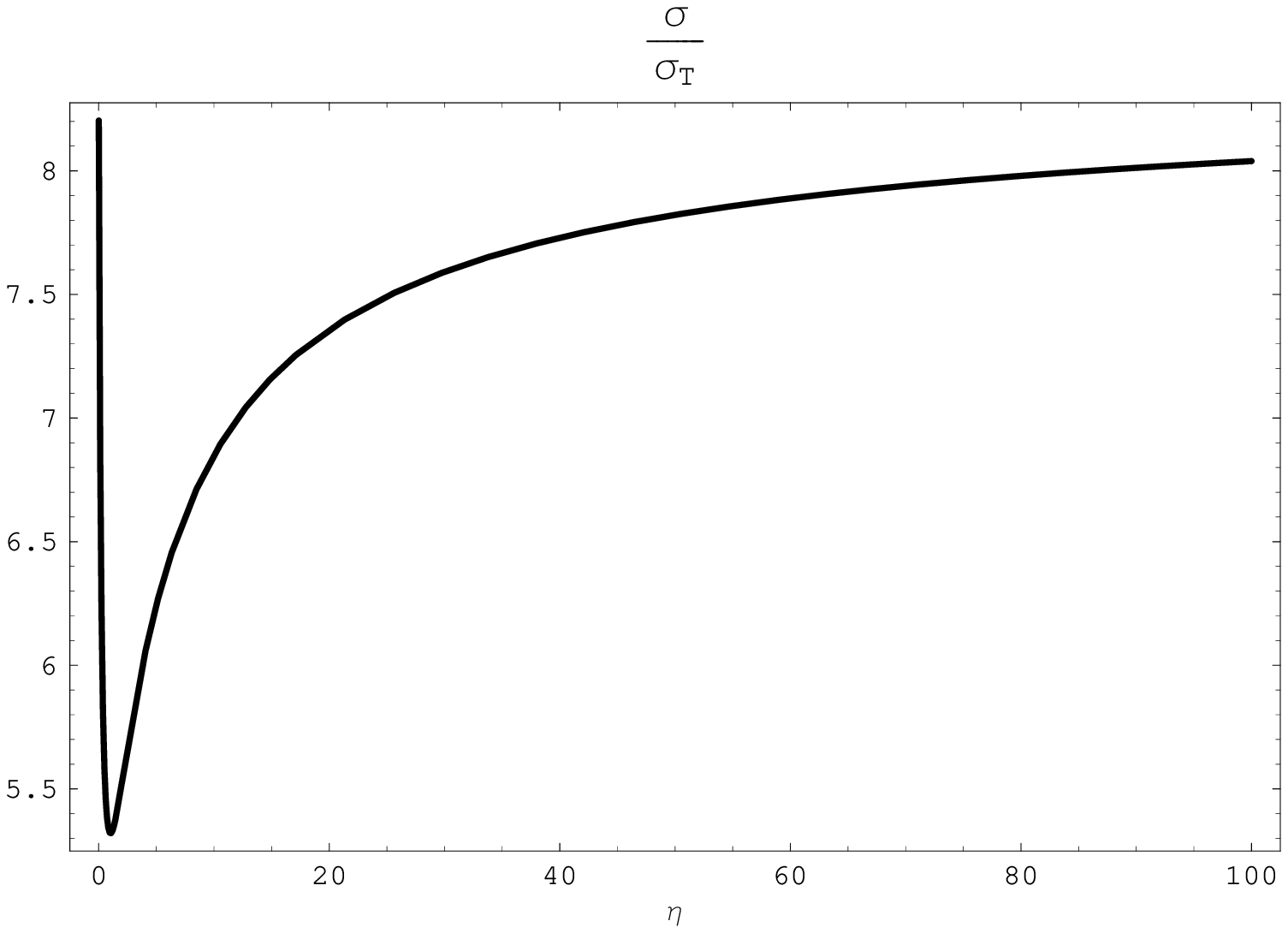}
\end{center}
\caption{Total cross section normalized to the Thompson value as a function of
the energy of the incident photon in the laboratory frame,
where $\eta=\omega/m$.}%
\label{sig}%
\end{figure}
\end{center}

These are precisely the parameter values that enter the description of the
electromagnetic properties of the $W$ boson in the Standard Model. However,
our results, being based on first principles, are valid for any elementary
spin-1 particle described by a four-vector field. 
Notice that in our derivation no assumptions have been made
about other interactions of the vector particle and its Abelian or non-Abelian
nature. In the following section we discuss our results in the context of
non-Abelian gauge theories. Before this, and for the sake of completeness, we
present our results for Compton scattering off any elementary massive vector 
particle in terms of the dimensionless variable $\eta$. 
The full angular distribution for the case $g=2$, $\xi=0$, is
\begin{align}
\frac{d\sigma(2,0)}{d\Omega}  &  =\frac{r_{0}^{2}}{6\left(  1+\eta
(1-x)\right)  ^{4}}\left[  3+6\,\eta+11\eta^{2}+8\eta^{3}+4\,\eta^{4}%
+x^{4}\,\eta^{2}\,\left(  3+4{\eta}^{2}\right)  \right. \label{ds20}\\
&  -2\,x^{3}\,\eta\,\left(  3+3\,\eta+4\,\eta^{2}+8\,\eta^{3}\right)
-2\,x\,\eta\,\left(  3+11\,\eta+12\,\eta^{2}+8\,\eta^{3}\right) \nonumber\\
&  \left.  +\,x^{2}\,\left(  3+6\,\eta+14\,\eta^{2}+24\,\eta^{3}+24\,\eta
^{4}\right)  \right]  .\nonumber
\end{align}
At high energies we obtain it flat according to
\begin{equation}
\left.  \frac{d\sigma(2,0)}{d\Omega}\right\vert _{\eta\gg1}=\frac{2}{3}%
r_{0}^{2},
\end{equation}
and the total cross section coincides with the Thompson one. Integrating
eq.~(\ref{ds20}) we obtain the total cross section as%
\begin{equation}
\sigma(2,0)=\frac{8\pi r_{0}^{2}}{3}\frac{2\,\eta\,\left(  9+54\,\eta
+129\,{\eta}^{2}+168\,{\eta}^{3}+140\,{\eta}^{4}+48\,{\eta}^{5}\right)
-3\,{\left(  1+2\,\eta\right)  }^{3}\,\left(  3+3\,\eta+4\,{\eta}^{2}\right)
\,\log(1+2\,\eta)}{12\,{\eta}^{3}\,{\left(  1+2\,\eta\right)  }^{3}}%
\end{equation}
In Fig.~\ref{dsig} we display the differential cross section as a function of
$x=\cos\theta$ for different values of the energy of the incident photon.
Starting from the classical angular distribution the radiation slightly peaks
in forward direction at intermediate energies but this effect is rapidly
damped and the angular distribution becomes flat at high energies.
In Fig.~\ref{sig} we show the total cross section as a function of
the energy of the incoming photon. It decreases from the Thompson value at
$\eta=0$ as the energy increases in the low energy region, 
reaches its minimum  
at $\eta=1$ and from there onward it smoothly rises approaching again
the Thompson value in the high energy limit, $\eta\gg1$.

\section{Partition of $g=2$ into Abelian and non-Abelian contributions for
non-Abelian vector gauge particles}

With the $g=2$ value for an elementary massive vector particle, the 
covariant projector formalism exploited
here has predicted for a second time a universal gyromagnetic ratio for a
high-spin particle. Earlier, same number has been obtained for spin-$3/2$ in
ref.~\cite{NKR}. There, and upon prohibiting spin-$3/2$ to spin-$1/2$
conversion, the electromagnetically gauged spin-$3/2$ Lagrangian was obtained
to depend on $g$ alone. The demand for causal spin-$3/2$ propagation within an
electromagnetic environment restricted then $g$ to $g=2 $. In this way the so
called Velo-Zwanziger problem of superluminal propagation of spin-$3/2$ fields
within an electromagnetic environment as suffered by the Rarita-Schwinger
formalism \cite{VZ32} was resolved and the solution related to the value of
the gyromagnetic factor. Also in this case, the nature of the particle,
Abelian versus non-Abelian, was irrelevant to the solution. On the other side,
it is well known that non-Abelian gauge theories contain "non-minimal"
electromagnetic interactions ( see \cite{Ferr} and references therein) which
contribute to the net electromagnetic couplings of gauge bosons. In order to
understand our results within the latter context let us consider the case of
the electroweak $W^{\pm}$ bosons for which the partition of its gyromagnetic
ratio into Abelian and non-Abelian contributions is well known \cite{BarryH}.
In that regard it is important to recall that the photon field, $A$, is not
among the four gauge bosons, $\mathbf{W}=\{W^{\pm},W^{3}\}$, and $B$, of the
$SU(2)_{L}\times U_{Y}(1)$ group but partakes both $B$ and $W^{3}$ according
to
\begin{align}
W_{\mu}^{3}  &  =\cos\,\theta_{W}Z_{\mu}^{0}+\sin\,\theta_{W}A_{\mu},
\nonumber\\
B_{\mu}  &  =\cos\,\theta_{W}A_{\mu}-\sin\,\theta_{W}Z_{\mu}^{0},
\label{WB}%
\end{align}
in standard notations. As long as the Abelian field $B$ belongs to $U(1)_{Y}$,
while the non-Abelian isovector field $\mathbf{W}$ is associated with
$SU(2)_{L}$, the Abelian contribution to the electromagnetic interactions of
the $W$ take their origin from $U(1)_{Y}$ gauging while the non-Abelian
ones arise from the $SU(2)_{L}$ gauging, both in combination with $e=g_{W}%
\sin\theta_{W}$ ($g_{W}$ stands for the universal electroweak coupling). The
three physical massive gauge bosons $W^{\pm}$, and $Z^{0}$ emerge only after
spontaneous $SU(2)_{L}\times U(1)_{Y}/U_{em}(1)$ breaking and it is only at 
this
level that one can identify their corresponding electromagnetic interactions.
What one observes is a partition of the gyromagnetic ratio in two sectors, the
non-Abelian one, $g_{na}=1$, as provided by the non-Abelian field tensor, and
the Abelian one, $g_{a}=2-g_{na}$, coming from the Abelian $U(1)_{Y}$ gauging.
The $g_{a}=1$ value required by the specifics of the electroweak gauge group
coincides by chance with the one provided by Proca theory, so that using
Proca's Lagrangian in the Standard Model is of no harm. However, for any other
gauge group, that provides a $g_{na}\not =1$, this concept will necessarily
collapse. {}For a general non-Abelian theory, based on a group, call it $G$,
and different from the electroweak one, $U(1)_{em}$ will manifest itself only
after the spontaneous $G/U_{em}(1)$ breaking at low energies. Concerning
physics beyond the Standard Model both aspects are completely unknown, so far.
Apparently, the respective $g_{na}$ value will depend both on the $U(1)_{em}$
embedding in $G$ and on the details of the spontaneous symmetry breaking and
can be lesser or bigger than $1$. It is obvious that Proca's theory is not
applicable to this case. Instead, one can make use of the Lagrangian with the
free $g$ parameter as defined by eqs.~(\ref{Gammag}) and (\ref{Lagg}) and
employ it, this time at the level \textit{before\/} the spontaneous symmetry
breaking. Fixing $g_{a}$ to $g_{a}=2-g_{na}$ guarantees $g=g_{a}+g_{na}=2$ 
for any 
needed partition of the net gyromagnetic ratio at the final stage.

\noindent What after all should be abundantly clear is that whatever the
unknown group $G$, the $U_{em}(1)$ embedding in it, or the mechanisms for the
spontaneous symmetry breaking might be, with the Lagrangian defined by
eqs.~(\ref{Gammag}) and (\ref{Lagg}) one can always end up with a net 
gyromagnetic ratio of $g_{a}+g_{na}=2$ for a vector gauge boson. At any rate, 
the three first principles mentioned above must be respected by the final 
form of the
interaction of the vector particle with the electromagnetic field and the
respective electromagnetic properties concluded here will always hold valid.

\section{Conclusions and perspectives}

In this work we studied the structure of the Lagrangian of an elementary
vector particle ( massive particle transforming in the 
$\left(\frac{1}{2},\frac{1}{2} \right)$ representation of the HLG ) interacting
with an electromagnetic field. The Lagrangian's
derivation was based on the three fundamental principles of (i) Poincar\'{e}
invariance of space-time, (ii) $U(1)_{em}$ gauge symmetry of electromagnetism,
and (iii) unitarity bounds for the Compton scattering cross section. The first
two principles lead to a general Lagrangian depending on
two free parameters, $g$, and $\xi$, both required in the definition of the
four electromagnetic multipoles characterizing a vector particle. Requiring
the total cross section for Compton scattering off a vector target to
respect the unitarity bounds in the high-energy limit allows to fix the free
parameters to $g=2$, $\xi=0$ and thereby to determine the tree level
electromagnetic properties of \textit{any} vector particle, be it Abelian or
non-Abelian. It must have a magnetic dipole
moment of $\mu=e/m$ (a gyromagnetic ratio of $g=2$), an electric quadrupole
moment of $Q=-e/m^{2}$, and vanishing electric dipole and magnetic quadrupole
moments at tree level. Modifications to this picture can arise only at
one-loop level either through higher order electromagnetic effects, or,
through electromagnetic corrections induced by interactions with other
particles. For gauge vector bosons the electromagnetic couplings are
partitioned into non-Abelian ($g_{na}$) and Abelian ($g_{a}$) contributions
obeying the restriction, $g=g_{a}+g_{na}=2$. The specific respective $g_{a}$,
and $g_{na}$ values depend on the gauge group $G$, the embedding of
$U(1)_{em}$ in it, and the details of the spontaneous symmetry breaking
$G\rightarrow U_{em}(1)$ at low energies.

The results obtained here are valid for elementary  vector particles, 
i.e., massive spin-$1$ particles 
transforming in the $(\frac{1}{2},\frac{1}{2})$ 
representation of the HLG as is the case of the $W$ boson. This is certainly
not the only possible HLG representation for the description of spin-$1$
though the one of the widest spread,
and  it would be interesting to check validity of the concepts 
presented here for spin-$1$ fields
transforming in other representations such as 
the totally  anti-symmetric second rank tensor, $ (1,0)\oplus(0,1)$, 
(considered in \cite{Gasser}),
or, the totally symmetric one, $(1,1)$ 
(considered, among others, in ~\cite{GRF}). 
Although we expect the calculation to evolve
similarly to the one presented here, the new problems need to be
worked out anew, a task that is beyond the scope 
of the present study.
Finally, another challenge for future research  would be to explore 
within the context of the covariant projector formalism  
the link between $g=2$ and the renormalizability 
of an effective field theory as found in ref.~\cite{Djukanovic}.

\begin{acknowledgments}
The work of M. Napsuciale was supported by
CONACyT-M\'{e}xico under project CONACyT-50471-F,  DGICYT contract number 
FIS2006-03438 and the Generalitat Valenciana. This research is part of the 
EU Integrated Infrastructure Initiative Hadron Physics Project under contract 
number RII3-CT-2004-506078. 
Work partly supported by CONACyT-M\'{e}xico under grant number
CB-2006-01/61286.
\end{acknowledgments}

\section{appendix}
The explicit expression for the invariant amplitude in the 
Compton scattering off a vector target is a bit cumbersome and given by  

\begin{align}
\mathcal{M} &  =-e^{2}\left\{  2\left(  \zeta\cdot\zeta^{\prime}\right)
\left[  \frac{\left(  p\cdot\varepsilon\right)  \left(  p^{\prime}%
\cdot\varepsilon^{\prime}\right)  }{p\cdot k}-\frac{\left(  p\cdot
\varepsilon^{\prime}\right)  \left(  p^{\prime}\cdot\varepsilon\right)
}{p\cdot k^{\prime}}-\varepsilon\cdot\varepsilon^{\prime}\right]  -g\left[
\left(  \varepsilon^{\prime}\cdot\left[  \zeta,\zeta^{\prime}\right]  \cdot
k^{\prime}\right)  \left(  \frac{p\cdot\varepsilon}{p\cdot k}-\frac{p^{\prime
}\cdot\varepsilon}{p\cdot k^{\prime}}\right)  \right.  \right.
\label{Amplitude}\\
&  -\left.  \left(  \varepsilon\cdot\left[  \zeta^{\prime},\zeta\right]  \cdot
k\right)  \left(  \frac{p\cdot\varepsilon^{\prime}}{p\cdot k^{\prime}}%
-\frac{p^{\prime}\cdot\varepsilon^{\prime}}{p\cdot k}\right)  \right]
-2\xi\left[  \frac{p\cdot\varepsilon\left\langle \zeta\zeta^{\prime
}\varepsilon^{\prime}k^{\prime}\right\rangle +p^{\prime}\cdot\varepsilon
^{\prime}\left\langle \zeta^{\prime}\zeta\varepsilon k\right\rangle }{p\cdot
k}+\frac{p\cdot\varepsilon^{\prime}\left\langle \zeta\zeta^{\prime}\varepsilon
k\right\rangle +p^{\prime}\cdot\varepsilon\left\langle \zeta^{\prime}%
\zeta\varepsilon^{\prime}k^{\prime}\right\rangle }{p\cdot k^{\prime}}\right]
\nonumber\\
&  +\frac{g\xi}{2}\left[  \frac{1}{p\cdot k}\left(  k\cdot\zeta\left\langle
\varepsilon\zeta^{\prime}\varepsilon^{\prime}k^{\prime}\right\rangle
-\varepsilon\cdot\zeta\left\langle k\zeta^{\prime}\varepsilon^{\prime
}k^{\prime}\right\rangle +k^{\prime}\cdot\zeta^{\prime}\left\langle
\varepsilon^{\prime}\zeta\varepsilon k\right\rangle -\zeta^{\prime}%
\cdot\varepsilon^{\prime}\left\langle k^{\prime}\zeta\varepsilon
k\right\rangle \right)  \right.  \nonumber\\
&  -\left.  \frac{1}{p\cdot k^{\prime}}\left(  k^{\prime}\cdot\zeta
\left\langle \varepsilon^{\prime}\zeta^{\prime}\varepsilon k\right\rangle
-\varepsilon^{\prime}\cdot\zeta\left\langle k^{\prime}\zeta^{\prime
}\varepsilon k\right\rangle +k\cdot\zeta^{\prime}\left\langle \varepsilon
\zeta\varepsilon^{\prime}k^{\prime}\right\rangle -\zeta^{\prime}%
\cdot\varepsilon\left\langle k\zeta\varepsilon^{\prime}k^{\prime}\right\rangle
\right)  \right]  \nonumber\\
&  +\frac{g^{2}}{2}\left[  \frac{1}{p\cdot k}\left[  k\cdot\zeta\left(
\varepsilon^{\prime}\cdot\left[  \varepsilon,\zeta^{\prime}\right]  \cdot
k^{\prime}\right)  -\varepsilon\cdot\zeta\left(  k\cdot\left[  \varepsilon
^{\prime},k^{\prime}\right]  \cdot\zeta^{\prime}\right)  \right]  -\frac
{1}{p\cdot k^{\prime}}\left[  k^{\prime}\cdot\zeta\left(  \varepsilon
\cdot\left[  \varepsilon^{\prime},\zeta^{\prime}\right]  \cdot k\right)
-\varepsilon^{\prime}\cdot\zeta\left(  k^{\prime}\cdot\left[  \varepsilon
,k\right]  \cdot\zeta^{\prime}\right)  \right]  \right]  \nonumber\\
&  +\frac{\xi^{2}}{2}\left.  \left[  \frac{1}{p\cdot k}\left\langle \mu
\zeta\varepsilon^{\prime}k^{\prime}\right\rangle \left\langle \mu\zeta
^{\prime}\varepsilon k\right\rangle -\frac{1}{p\cdot k^{\prime}}\left\langle
\mu\zeta\varepsilon^{\prime}k^{\prime}\right\rangle \left\langle \mu
\zeta^{\prime}\varepsilon k\right\rangle \right]  \right\}  \nonumber\\
&  +\frac{e^{2}}{2m^{2}}\left\{  \left(  g-2\right)  \xi\left[  \frac
{1}{p\cdot k^{\prime}}\left[  \left(  \varepsilon\cdot\left[  p,\zeta\right]
\cdot k\right)  \left\langle p^{\prime}\zeta^{\prime}\varepsilon^{\prime
}k^{\prime}\right\rangle +\left(  \varepsilon^{\prime}\cdot\left[  p^{\prime
},\zeta^{\prime}\right]  \cdot k^{\prime}\right)  \left\langle p\zeta
\varepsilon k\right\rangle \right]  \right.  \right.  \nonumber\\
&  -\left.  \frac{1}{p\cdot k^{\prime}}\left[  \left(  \varepsilon^{\prime
}\cdot\left[  p,\zeta\right]  \cdot k^{\prime}\right)  \left\langle p^{\prime
}\zeta^{\prime}\varepsilon k\right\rangle +\left(  \varepsilon\cdot\left[
p^{\prime},\zeta^{\prime}\right]  \cdot k\right)  \left\langle p\zeta
\varepsilon^{\prime}k^{\prime}\right\rangle \right]  \right]  +\left(
g-2\right)  ^{2}\left[  \frac{1}{p\cdot k}\left(  \varepsilon^{\prime}%
\cdot\left[  p^{\prime},\zeta^{\prime}\right]  \cdot k^{\prime}\right)
\left(  \varepsilon^{\prime}\cdot\left[  p^{\prime},\zeta^{\prime}\right]
\cdot k^{\prime}\right)  \right.  \nonumber\\
&  -\left.  \frac{1}{p\cdot k^{\prime}}\left(  \varepsilon\cdot\left[
p^{\prime},\zeta^{\prime}\right]  \cdot k\right)  \left(  \varepsilon
\cdot\left[  p^{\prime},\zeta^{\prime}\right]  \cdot k\right)  \right]
+\left.  \xi^{2}\left[  \frac{1}{p\cdot k}\left\langle p\zeta\varepsilon
k\right\rangle \left\langle p^{\prime}\zeta^{\prime}\varepsilon^{\prime
}k^{\prime}\right\rangle -\frac{1}{p\cdot k^{\prime}}\left\langle
p\zeta\varepsilon^{\prime}k^{\prime}\right\rangle \left\langle p^{\prime}%
\zeta^{\prime}\varepsilon k\right\rangle \right]  \right\}. \nonumber
\end{align}
Here we used Holstein's notation \cite{BarryH}%
\begin{equation}
S\cdot\left[  Q,R\right]  \cdot T=S\cdot QR\cdot T-S\cdot RQ\cdot T ,
\end{equation}
and defined%
\begin{equation}
\left\langle \alpha ABC\right\rangle \left\langle \alpha A^{\prime}B^{\prime
}C^{\prime}\right\rangle =\epsilon_{\alpha\beta\mu\nu}A^{\beta}B^{\mu}C^{\nu
}\epsilon_{\quad\sigma\rho\gamma}^{\alpha}A^{\prime\beta}B^{\prime\mu
}C^{\prime\nu}.
\end{equation}
The resulting cross section is then obtained as
\begin{align}
\frac{d\sigma(g,\xi)}{d\Omega} &  =\frac{r_{0}^{2}}{96\,\left(  1+\eta
(1-x)\right)  ^{4}}\left\{  48+96\,\eta+x^{4}\,\eta^{2}\,\left(  48+\eta
^{2}\,\left(  -4+4\,g+g^{2}+\xi^{2}\right)  ^{2}\right)  \right.  \\
&  +2\,\eta^{3}\,\left(  80-48\,g+3\,g^{4}+24\,\xi^{2}+3\,\xi^{4}%
+g^{2}\,\left(  8+6\,\xi^{2}\right)  \right)  \nonumber\\
&  +2\,\eta^{2}\,\left(  104-48\,g+3\,g^{4}+24\,\xi^{2}+3\,\xi^{4}%
+g^{2}\,\left(  8+6\,\xi^{2}\right)  \right)  \nonumber\\
&  +\eta^{4}\,\left(  80-80\,g^{3}+21\,g^{4}+88\,\xi^{2}+21\,\xi
^{4}-16\,g\,\left(  8+5\,\xi^{2}\right)  +2\,g^{2}\,\left(  68+21\,\xi
^{2}\right)  \right)  \nonumber\\
&  -2\,x^{3}\,\eta\,\left[  48+48\,\eta-\eta^{2}\,\left(  -16+48\,g+g^{4}%
+8\,\xi^{2}+\xi^{4}+2\,g^{2}\,\left(  -20+\xi^{2}\right)  \right)  \right.
\nonumber\\
&  +\left.  \eta^{3}\,\left(  16+12\,g^{3}+5\,g^{4}-8\,\xi^{2}+5\,\xi
^{4}+4\,g\,\left(  -4+3\,\xi^{2}\right)  +2\,g^{2}\,\left(  -4+5\,\xi
^{2}\right)  \right)  \right]  \nonumber\\
&  +2\,x^{2}\,\left[  24+48\,\eta-\eta^{2}\,\left(  -64+48\,g+g^{4}+8\,\xi
^{2}+\xi^{4}+2\,g^{2}\,\left(  -20+\xi^{2}\right)  \right)  \right.
\nonumber\\
&  \left.  -\eta^{3}\,\left(  -80-16\,g^{3}+g^{4}+8\,\xi^{2}+\xi^{4}%
+2\,g^{2}\,\left(  -20+\xi^{2}\right)  -16\,g\,\left(  -5+\xi^{2}\right)
\right)  \right.  \nonumber\\
&  \left.  +\eta^{4}\,\left(  48-28\,g^{3}+19\,g^{4}+40\,\xi^{2}+19\,\xi
^{4}-4\,g\,\left(  12+7\,\xi^{2}\right)  +g^{2}\,\left(  40+38\,\xi
^{2}\right)  \right)  \right]  \nonumber\\
&  -2\,x\,\eta\,\left[  48+16\,\eta\,\left(  7+g^{3}+g\,\left(  -2+\xi
^{2}\right)  \right)  \right.  \nonumber\\
&  +\left.  \eta^{3}\,\left(  80-76\,g^{3}+25\,g^{4}+88\,\xi^{2}+25\,\xi
^{4}+10\,g^{2}\,\left(  12+5\,\xi^{2}\right)  -4\,g\,\left(  28+19\,\xi
^{2}\right)  \right)  \right.  \nonumber\\
&  \left.  +\eta^{2}\,\left(  16\,g^{3}+3\,g^{4}+16\,g\,\left(  -5+\xi
^{2}\right)  +g^{2}\,\left(  8+6\,\xi^{2}\right)  +3\,\left(  48+8\,\xi
^{2}+\xi^{4}\right)  \right)  \right]  \left.  {}\right\},
\label{dsgt}%
\end{align}
where $r_{0}=\frac{\alpha}{m}$ stands for the classical radius, $\eta
=\omega/m$, $x=\cos\theta$, and we used
\begin{equation}
\omega^{\prime}=\frac{\omega}{\left(  m+\omega(1-x)\right)  },
\end{equation}
for the energy of the final photon $\omega^{\prime}$ .

\end{document}